\documentclass{emulateapj}
\usepackage{natbib}
\usepackage{graphicx}
\usepackage{amssymb}

\def\msun{{~M}_{\odot}}

\def\be{\begin{equation}}
\def\ee{\end{equation}}
\def\ergs{{\rm\,erg\,s^{-1}}}

\catcode`\@=11 
\def\@versim#1#2{\vcenter{\offinterlineskip
        \ialign{$\m@th#1\hfil##\hfil$\crcr#2\crcr\sim\crcr } }}
\def\mpy{M_{\sun} \ {\rm yr^{-1}}}

\begin{document}

\shorttitle{A large dynamical range and the density profile of hot accretion flows}
\shortauthors{Yuan, Wu \& Bu}

\title{Numerical Simulation of Hot Accretion Flows (I): A Large Radial Dynamical Range and the Density Profile of Accretion Flow}

\author{Feng Yuan, Maochun Wu and Defu Bu}
\affil{Key Laboratory for Research in Galaxies and Cosmology, Shanghai Astronomical Observatory, Chinese Academy of Sciences, 80 Nandan Road, Shanghai 200030, China;
fyuan,mcwu,dfbu@shao.ac.cn}

\begin{abstract}
Numerical simulations of hot accretion flow, both hydrodynamical and magnetohydrodynamical, have shown that the mass accretion rate decreases with decreasing radius; consequently the density profile of accretion flow becomes flatter compared to the case of a constant accretion rate. This result has important theoretical and observational implications. However, because of technical difficulties, the radial dynamic range in almost all previous simulations usually spans at most two orders of magnitude. This small dynamical range, combined with the effects of boundary conditions, makes the simulation results suspectable. Especially, the radial profiles of density and inflow rate may not be precise enough to be used to compare with observations. In this paper we present a ``two-zone'' approach to expand the radial dynamical range
from two to four orders of magnitude. We confirm previous results and find that from $r_s$ to $ 10^4r_s$ the radial profiles of accretion rate and density can be well described by $\dot{M}(r)\propto r^s$ and $\rho\propto r^{-p}$.
The values of (s, p) are (0.48, 0.65) and (0.4, 0.85), for viscous
parameter $\alpha=0.001$ and $0.01$, respectively. Or more precisely, the accretion rate is constant (i.e., $s=0$) within $\sim 10r_s$; but beyond $10r_s$, we have $s=0.65$ and $0.54$ for $\alpha=0.001$ and $0.01$, respectively. We  find that the values of both $s$ and $p$ are similar in all numerical simulation works, including previous and the present ones, no matter a magnetic field is included or not and what kind of initial conditions are adopted. Such an apparently surprising ``common'' result can be explained by the most updated version of the adiabatic inflow-outflow model (ADIOS). The density profile we obtain is in good quantitative agreement with that obtained from the detailed observations and modeling to Sgr A* and NGC~3115. The origin and implication of such a profile will be investigated in a subsequent paper.
\end{abstract}


\section{INTRODUCTION}

Hot accretion flow, such as advection-dominated accretion flows
(ADAFs; Ichimura 1977; Rees et al. 1982; Narayan \& Yi 1994; 1995; Abramowicz et al. 1995; see
Narayan, Mahadevan \& Quataert 1998; Kato, Fukue \& Mineshige 1998; and Yuan \& Narayan 2013 for reviews), is of great interest because of its widespread applications in low-luminosity AGNs,
including the spuermassive black hole in our Galactic center, and
the quiescent and hard states of black hole X-ray binaries (see
reviews by Narayan 2005; Yuan 2007; Narayan \& McClintock 2008; Ho 2008; and Yuan 2011). In the early analytical works, the mass accretion rate is assumed to be independent of radius, $\dot{M}(r) = {\rm constant}$. In this case, the radial profile of density satisfies $\rho (r)\propto r^{-3/2}$ (e.g., Narayan \& Yi 1994). Numerous hydrodynamical (HD) and magnehydrodynamical (MHD) numerical simulations have been done, with most of them focusing on the dynamics of the accretion flow (e.g., Igumenshchev \& Abramowicz 1999, 2000; Stone, Pringle \& Begelman 1999, hereafter SPB99; Stone \& Pringle 2001; Hawley, Balbus \& Stone 2001; Machida, Matsumoto \& Mineshige 2001; McKinney \& Gammie 2002; Hawley \& Balbus 2002; Igumenshchev, Narayan \& Abramowicz 2003; Pen, Matzener \& Wong 2003; De Villiers, Hawley \& Krolik 2003;  Proga \& Begelman 2003a, 2003b; Pang et al. 2011; McKinney, Tchekhovskoy \& Blandford 2012; Narayan et al. 2012). The effect of strong radiation was studied by Yuan \& Bu (2010), and most recently by Li, Ostriker \& Sunyaev (2012). One of the most surprising -- which is perhaps also the most important -- findings of these simulations is that the mass accretion rate (or precisely the inflow rate; refer to eq. [7] in the present paper for its definition), is found to be not a constant; but rather, it decreases with decreasing radius. Denoting the mass accretion rate as $\dot{M}(r)\propto r^{s}$, numerical simulations have found that $s\sim 0.5-1$ (see \S 4.1 for a review). Consequently, the density profile flattens compared to the previous $\rho(r)\propto r^{-1.5}$: we now have $\rho(r)\propto r^{-p}$ with $p\la 1$. Such results have obtained strong observational supports in the case of the supermassive black hole in our Galactic center, Sgr A* (Yuan, Quataert \& Narayan 2003; refer to \S4.2.1 for details.), the low-luminosity AGN NGC~3115 (Wong et al. 2011; refer to \S4.2.2 for details), and black holes in elliptical galaxies (Di Matteo et al. 2000; Mushotzky et al. 2000).

In addition to the obvious theoretical interest, the radial profiles of accretion rate and density have also important
observational applications. This is because they will determine the emitted spectrum and other radiative features of an accretion flow (e.g., Quataert \& Narayan 1999; Yuan, Quataert \& Narayan 2003). For example, in the case of Sgr A*, the mass accretion rate at the Bondi radius can be determined directly by observations. Then depending on
the radial profile of the accretion rate (or more exactly the density), the radiation of the accretion flow is completely different. The radial profile of accretion
rate also determines the evolution of black hole mass and spin. In
many current numerical simulations, due to resolution difficulty, we
can at most resolve the Bondi radius and determine the Bondi
accretion rate there. Then the evolution of mass and spin of black holes
will be determined by the fraction of the Bondi accretion rate
that finally falls onto the black hole, which is determined by the
radial profile of accretion rate.

It is thus important to carefully investigate the radial profiles of
accretion rate and density. One problem of almost all previous simulations is that the radial dynamical range is rather small, usually at most two orders of magnitude. This is technically because it would be so time-consuming that it is almost impossible to simulate an accretion flow, if the dynamical range is too large. In addition, as is well known the simulation results are usually not reliable close to
the boundary due to the boundary condition effects. These cast some shade on the previous simulation results, especially the exact quantitative radial profiles of the physical quantities. The situation is even worse for MHD simulations compared to HD ones. Because the Alfv\'en speed is very large in the region of low density and strong magnetic field, MHD simulation is much more expensive than HD simulations. The radial dynamical range is thus more limited, and it is harder to evolve the
simulation for long time which further constraints the range over
which a steady state is reached.

A large dynamical range is also useful to investigate
the following problem. Previous works (SPB99; Igumenshchev \&
Abramowicz 1999; Yuan \& Bu 2010) have found that the Bernoulli
parameter of most outflow in their simulations is indeed negative\footnote{In our subsequent paper (Yuan, Bu \& Wu 2012, hereafter Paper II) we found that depending on the initial condition of the simulation, the Bernoulli parameter can be negative or positive.}. One may expect that outflow may not be able to escape to infinity, but may rejoin the accretion flow at a certain distance. The fact that all previous simulations never found the accumulation of matter in the accretion flow could be because of two possible reasons. One is that the simulation time is not long enough for accumulation to occur. However, the radial velocity of outflow is roughly in the order of the local Keplerian velocity (Paper II). We can therefore expect that if the outflow would finally rejoin the accretion flow, the timescale required for accumulation would be shorter than the accretion timescale by a factor of $\alpha$. { Thus this possibility is unlikely.} Another possible reason is that the radial dynamical range is too small. The accumulation may occur at a radius larger than the outer boundary of all current simulations. To examine this possibility, a larger dynamical range is required.

In the present paper we simulate two-dimensional HD accretion flows. Especially, a ``two-zone'' approach { will be} adopted which helps us to overcome the technical problem and achieve a large dynamic range spanning four orders of magnitude.  It is widely believed that in reality the magnetic stress associated with the MHD turbulence driven by the magnetorotational instability (MRI) transfers angular momentum (Balbus \& Hawley 1991; 1998). Therefore, we should in principle use MHD simulation. In the present paper we don't include magnetic field but instead include an anomalous shear stress to mimic the magnetic stress. However, as we will describe in Paper II, in HD and MHD accretion flows, the mechanisms of producing the accretion rate profile are different. One may ask whether the radial profile of accretion rate obtained in the present HD simulation is the same with that obtained in more realistic MHD simulations. In this regard, the recent work by Begelman (2012) gives a positive answer. The comparison between our HD simulation with MHD simulations also confirms this point (refer to \S4.1 for details).

The structure of the present paper is as follows. In \S2, we introduce the details of our ``two-zone'' simulation approach. The results of simulations are presented in \S3. We find that the profiles of the mass accretion rate and density
still follows a perfect power-law form and the power-law index almost remains unchanged compared to previous results. In \S4 we compare our results with previous HD and MHD simulation works (\S4.1) and observations (\S4.2). We find broadly good agreement among them. The last section (\S5) devotes to a summary.

\section{Method}

\subsection{Equations}
The equations describing the hydrodynamics of accretion flows are
taken from SPB99:
\begin{equation}
\frac{d\rho}{dt}+\rho\nabla\cdot \mathbf{v}=0,\label{cont}
\end{equation}
\begin{equation}
\rho\frac{d\mathbf{v}}{dt}=-\nabla p-\rho\nabla
\psi+\nabla\cdot\mathbf{T}, \label{rmon}
\end{equation}
\begin{equation}
\rho\frac{d(e/\rho)}{dt}=-p\nabla\cdot\mathbf{v}+\mathbf{T}^2/\mu.
\label{rmon}
\end{equation}
We use the spherical coordinates $(r, \theta, \phi)$ to solve these
equations. In the above equations, $\rho$, $p$, $v$, and $e$  are
the mass density, pressure, velocity, and internal energy density,
respectively. Here $d/dt\equiv
\partial /\partial t + \mathbf{v}\cdot \nabla$. We adopt an
adiabatic equation of state $P=(\gamma -1)e$ with $\gamma =5/3$.
$\psi$ is the gravitational potential. We adopt the Paczy\'nsky \&
Wiita (1980) potential $\psi=-GM/(r-r_s)$, where $M$ is the center black
hole mass, $G$ is the gravitational constant and $r_s \equiv 2GM/c^2$. The self gravity of the accretion flow is neglected.

In a real accretion flow, the Maxwell stress associated with the MHD
turbulence driven by the magnetorotational instability transfers the
angular momentum. In the present work, we do not include magnetic field; instead we follow SPB99 and add the final terms in equations (2) and (3) to represent the anomalous shear stress and
the corresponding heating. Here $\mathbf{T}$ is anomalous stress tensor. To
approximate the effect of magnetic stress, again following SPB99, we
assume that only azimuthal components of the stress are non-zero
because MRI is driven only by the shear associated with the orbital
dynamics:
\begin{equation}
  T_{r\phi} = \mu r \frac{\partial}{\partial r}
    \left( \frac{v_{\phi}}{r} \right),
\end{equation}
\begin{equation}
  T_{\theta\phi} = \frac{\mu \sin \theta}{r} \frac{\partial}{\partial
  \theta} \left( \frac{v_{\phi}}{\sin \theta} \right) .
\end{equation}
This approximation is supported by three-dimensional MHD simulations
(e.g., Hawley, Gammie \& Balbus 1996; Stone et al. 1996). We
emphasize this approximation because the inclusion of other
components was found to produce quite different results
(Igumenshchev \& Abramowicz 1999). The viscosity coefficient
$\mu=\nu\rho$ determines the magnitude of the stress and $\nu$ is
the kinematic viscosity coefficient. We adopt the form $\nu=\alpha
{r^{1/2}}$ because it corresponds to the usual ``$\alpha$''
description (SPB99). In the present paper we investigate two cases,
i.e., $\alpha$=0.001 (Model A) and $\alpha$=0.01 (Model B).  We
choose units such that $M=G=1$ and $c=10\sqrt{2}$.

\subsection{Two-zone approach}

We aim at  simulations with a large dynamical range of four orders of
magnitude in radius. Directly simulating such a large range is technically almost impossible. For this purpose, we divide the whole
simulation domain into two zones. The inner zone is from 1.35 $r_s$
to $200 r_s$ while the outer one from 100 $r_s$ to 40000 $r_s$. We
simulate these two zones separately, following the four steps
described below. For each zone, we use the ZEUS-2D code (Stone \&
Norman 1992) in spherical geometry to solve the equations. The two
modifications to the code are to include the shear stress terms and
the implementation of the Paczy\'nsky \& Wiita (1980) potential. We
adopt the same non-uniform grid in both the radial and angular
directions as in SPB99. The resolution is $N_r =168$ and $N_\theta
=88$.

{\em Step one}. We first simulate the outer zone. Following SPB99,
the initial state of our simulation is an equilibrium torus with a
constant specific angular momentum. The readers are referred to SPB99 for the  description of the torus. We set the radius of the
maximum density at $10000r_s$, the maximum density of the
torus $\rho_{\rm max}=1.0$, the density of the medium $\rho_0=10^{-4}$,
and pressure $ p_0=\rho_0/r$. The standard outflow boundary
conditions (projection of all dynamical variables) at both the inner
and outer boundaries are adopted.

{\em Step two.} We then simulate the inner zone. We inject the gas
at the outer boundary. The values of density, specific internal
energy, and velocity of the injected flow are taken from the
steady simulation results of Step one at $200r_s$. We do not use $100r_s$ because we hope to avoid the effect of the inner boundary. We adopt the outflow boundary condition at the inner boundary.

{\em Step three.} Obviously, simply connecting the simulation
results of the last two steps is generally not self-consistent. This
is because, for example, the inner zone will obviously produce some outflow at the outer boundary and these outflow will be injected into the outer
zone. This effect has not been taken into account in Step one. We
therefore simulate the outer zone again, but this time we use the
results of Step two at $100 r_s$ as the inner boundary condition. Again, here we do not choose $200 r_s$ because we hope to avoid the effect of outer boundary. In this way, we can capture all the outflow produced from the inner region and preserve all their properties such as their velocity and
internal energy, and thus their (negative) Bernoulli
parameter. Thus, we should be able to observe whether these outflow will
rejoin the accretion flow and accumulate somewhere in the outer
zone. After the outer zone reaches the steady state, we simulate the inner zone again, following the method described in Step two.

{\em Step four.} We plot the radial profiles of various physical quantities along different $\theta$ of the inner and outer zones. We by-eye look at the curves in each zone to judge the convergence, i.e., whether the results in each zone obtained in two adjacent steps are consistent with each other. If not, we do iteration to improve. Usually we can get satisfactory convergence after up to three iterations.

However, we note that this iteration approach does not guarantee a complete consistency between the two zones. In principle, we should transfer all information of the fluid at each time step from one zone to another. We assume steady solution is reached and only transfer part of the information at one snapshot. However, our approach does transfer some information between the two zones, and these information is perhaps the most important for the dynamics of accretion flow. Using this ``two-zone'' approach, we can easily simulate the accretion flow
with a dynamic range spanning four orders of magnitude in radius.
Compared to the usual ``one-zone'' approach, the calculation time
costed is about $10^3/N\sim$ several hundreds times shorter for our problem (here $N$ is the number of iterations). This approach could potentially be used in other simulation problems which require a large dynamical range. But we should emphasize that the underlying assumption of this approach is that steady solutions
exist. This is satisfied for our problem.

\section{Results}

\subsection{Model A: the case of $\alpha=0.001$}

\begin{figure*}
\hspace{1cm}\includegraphics[width=15.cm]{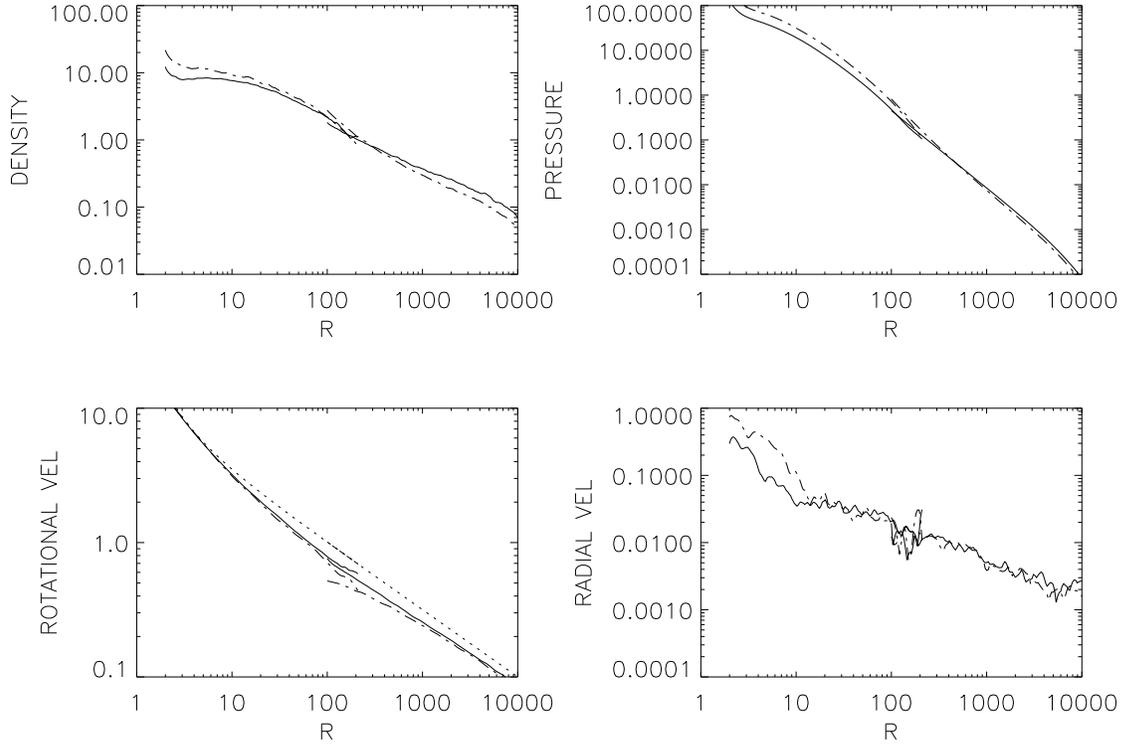}
\caption{Radial structure of the accretion flow (two zones) of Model
A after iteration. All quantities have been averaged over time and
the polar angle between $\theta=84^{\circ} $ and
$\theta=96^{\circ}$. The dashed line in the plot of rotational
velocity $v_{\phi}$ denotes Keplerian rotation at the equator. For
comparison, the results of {\em Step one} and {\em Step two} are shown by the dot-dashed lines.}
\end{figure*}

\begin{figure*}
\hspace{1.5cm}\includegraphics[width=14cm]{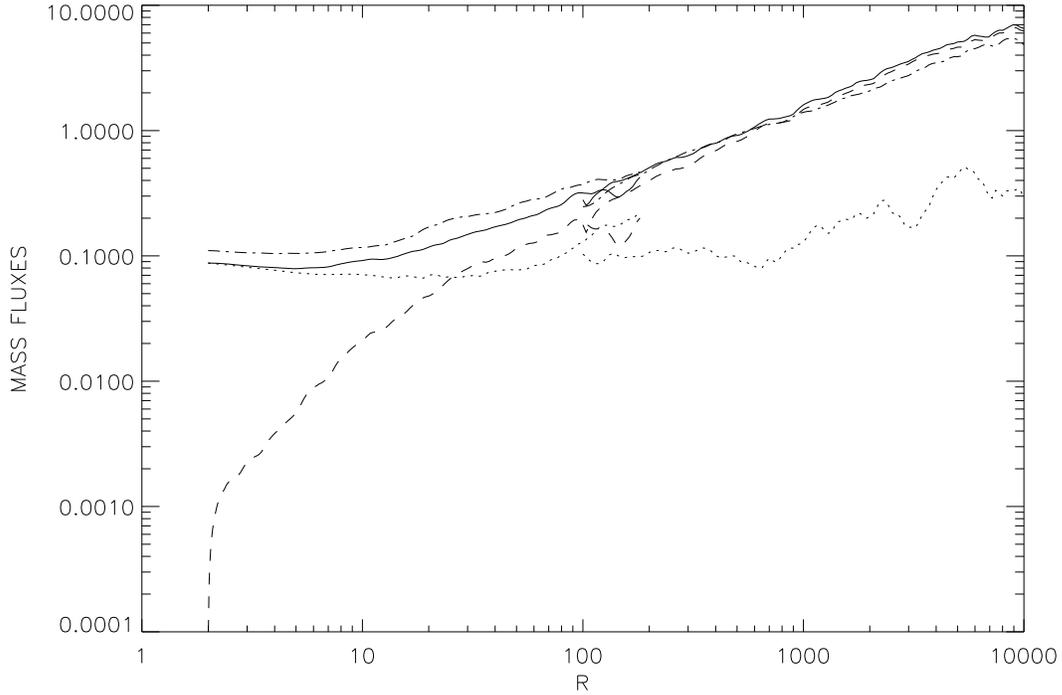}
\caption{Time-averaged and angle integrated mass accretion rate of
Model A. The solid, dashed, and dotted lines are for the inflow rate
$\dot{M}_{\rm in}$,  outflow rate $\dot{M}_{\rm out}$, and net rate
$\dot{M}_{\rm acc}$, respectively. For comparison, the results of
inflow rate of {\em Step one} and {\em Step two} are shown by the dot-dashed
line.} \label{fig:inflowoutflow}
\end{figure*}

Figure 1 shows the time-averaged radial distribution of physical quantities near the
equator of the whole region of the accretion flow (both the inner
and outer zones). The results are averaged over time and angle
between $\theta=84^{\circ} $ and $\theta=96^{\circ}$. The solid
lines are the final results of {\em Step four}. For comparison aim we also show by the dot-dashed lines the results of {\em Step one} and {\em Step two}. We can see that the profiles slightly become flatter after convergence is achieved. The density, gas pressure, rotation velocity and radial velocity can be described by a power law scaling with radius, \be \rho \propto r^{-0.65}, \hspace{0.5cm} P\propto r^{-1.7}, \hspace{0.5cm}v_{\phi} \propto r^{-0.5}, \hspace{0.5cm}~v_r \propto r^{-0.55}. \label{radialscaling}\ee  These scalings are measured away from the innermost region, i.e., $r\ga 10r_s$. We avoid the region within $10r_s$ because of two reasons. Firstly, as emphasized by Narayan et al. (2012) and also confirmed by the radial velocity plot in Figure 1, close to the black hole, the radial velocity of the flow increases inward much more rapidly because of the strong gravity, thus it deviates from its scaling extrapolated from the region of $r\ga 10r_s$. Secondly, as we will see from Figure 2 below, within $\sim 10r_s$, the inflow rate is a constant and there is little outflow. Therefore, in this sense this region is also ``special'' compared to the region outside $\sim 10r_s$. We note that these two ``inner boundary effects'' do not exit if a Newtonian potential is adopted.

These results are roughly consistent with SPB99, where they found $\rho\propto r^{-0.5}$, $P\propto r^{-1.5}$ and $v_r\propto r^{-0.5}$. We have done some test calculations and found that the small discrepancy is because we adopt the Paczy\'nsky \&
Wiita potential while a Newtonian potential is adopted in SPB99. The scaling of the radial and rotational velocities are also consistent with the self-similar solution of ADAF (Narayan \& Yi 1994)\footnote{ADIOS has the same scaling with ADAF for the radial and rotational velocities. ADIOS differs from ADAF only on the scaling of density and accretion rate. See eqs. (14) and (16) in Narayan et al. (2012).}, where we have $v_r\propto r^{-0.5}, v_{\phi}\propto r^{-0.5}$. The deviations of $\rho$ and $P$ from the self-similar solution are significant. This is because of the inward decrease of mass accretion rate, as we will describe below. We note that the scaling we obtain should be more reliable than previous works because of our extremely large radial dynamical range.

In Figure 2, we plot the time-averaged and angle-integrated mass
accretion rate of the whole region. Following SPB99, the mass inflow
and outflow rates, $\dot {M}_{\rm in}$ and $\dot {M}_{\rm out}$, are
defined as follows,
\begin{equation}
 \dot{M}_{\rm in}(r) = 2\pi r^{2} \int_{0}^{\pi} \rho \min(v_{r},0)
   \sin \theta d\theta,
\end{equation}
\begin{equation}
 \dot{M}_{\rm out}(r) = 2\pi r^{2} \int_{0}^{\pi} \rho \max(v_{r},0)
    \sin \theta d\theta.
\end{equation}
The net mass accretion rate is defined as,
\begin{equation}
\dot{M}_{\rm acc}(r)=\dot{M}_{\rm in}(r)+\dot{M}_{\rm out}(r).
\end{equation}
The rates of inflow and outflow, and the net rate are denoted by the solid,
dashed, and dotted lines, respectively. Note that the results are obtained by time-average the integral rather than integrating the time-averages. Also shown in the figure by the dot-dashed line is the inflow rate obtained in {\em Step one} and {\em Step two}. We see that the final curves of inflow rate after {\em Step four} becomes steeper  compared to the results of {\em Step one} and {\em Step two}. The net rate
close to the outer boundary of the outer zone is not constant,  indicating that it
requires longer time to get the fully steady solution there. This is
because the viscosity coefficient is very small so the accretion
timescale is very long. The radial profile of the inflow rate from
$\sim 2r_s$ to $10^4 r_s$ can be described by \be\dot{M}_{\rm in}(r)
=\dot{M}_{\rm in}(r_{\rm out})\left(\frac{r}{r_{\rm
out}}\right)^{0.48}.\label{inflowprofile}\ee   Or more precisely by, \be \dot{M}_{\rm in}(r)= \dot{M}_{\rm in}(r_{\rm out})\left(\frac{r}{r_{\rm
out}}\right)^{0.65}\label{inflowprofile2}\ee from
$\sim 10r_s$ to $10^4 r_s$, while it is almost constant within
$10r_s$. This is different
from SPB99. In that work, they found the inflow rate keeps
decreasing inward until the inner boundary, and the slope is steeper than ours, which is $\dot{M}_{\rm in} \propto r^{0.75}$ (``Run K'' in SPB99). Again, the reason is because we adopt a different gravitational potential of the black hole. Actually, we can see from the figure that, beyond $100r_s$ where the two types of potential is basically identical, the power-law index is $\sim 0.75$, in good consistency with SPB99. The deeper potential adopted in our work suppresses convection to some degree and makes the outflow weaker. In Paper II, we will analyze the origin of the inward decrease of the accretion rate. We find that  it is because of the mass loss in the outflow, and the the production of outflow is because of the gradient of gas pressure, i.e, the buoyant force. When the gravitational force is stronger, the gas pressure force plays a relatively minor role. In other words, in the innermost region, due to the strong gravity the accretion timescale is shorter  than the timescale required for the formation of outflow thus few outflow is produced.

If the flat density profile of $\rho\propto r^{-0.65}$  is because of the inward decrease of mass accretion rate, we should expect that the power-law index $p$ ($\rho\propto r^{-p}$) should be related to $s$ ($\dot{M}\propto r^{s}$) by $p=1.5-s$. From equations (\ref{radialscaling}) and (\ref{inflowprofile2}),  we have $p=0.65 <1.5-s=0.85$. The small deviation is mainly because the density profile depends not only on the inflow profile, but on the sum of the inflow and outflow rates, i.e., $\dot{M}_{\rm in}(r)-\dot{M}_{\rm out}(r)$. The radial profile of outflow rate is steeper than that of the inflow rate, \be\dot{M}_{\rm out}(r)\propto r^{0.87}\ee throughout the radius or \be\dot{M}_{\rm out}(r)\propto r^{0.8}\ee for the region beyond $r=100$. At large radii, $r\ga 100r_s$, the profiles of inflow and outflow rates are almost identical thus the relation $p=1.5-s$ is better satisfied, where $p \sim 0.65$ and $s \sim 0.75-0.8$.

The most notable  result from Figures 1 \& 2 is
that the profiles of physical quantities, such as density and mass
inflow rate, are well described by a single power-law form,
extending from $10^4 r_s$ to $\sim 2 r_s$. This confirms previous simulation results such as SPB99, although in those simulations the dynamical
range of the whole simulation domain usually spans only two orders of magnitude, and the steady solution is only reached within one order of magnitude because of the effect of outer boundary condition. This result indicates that
although the Bernoulli parameter of most of the mass outflow is
negative (SPB99 and Yuan \& Bu 2010), the outflow does not rejoin the accretion
flow and accumulate somewhere within $10^4r_s$. Physically, this
is likely because the flow is viscous thus there is an associated energy flux.

\subsection{Model B: the case of $\alpha=0.01$}

\begin{figure*}
\hspace{1cm}\includegraphics[width=15.cm]{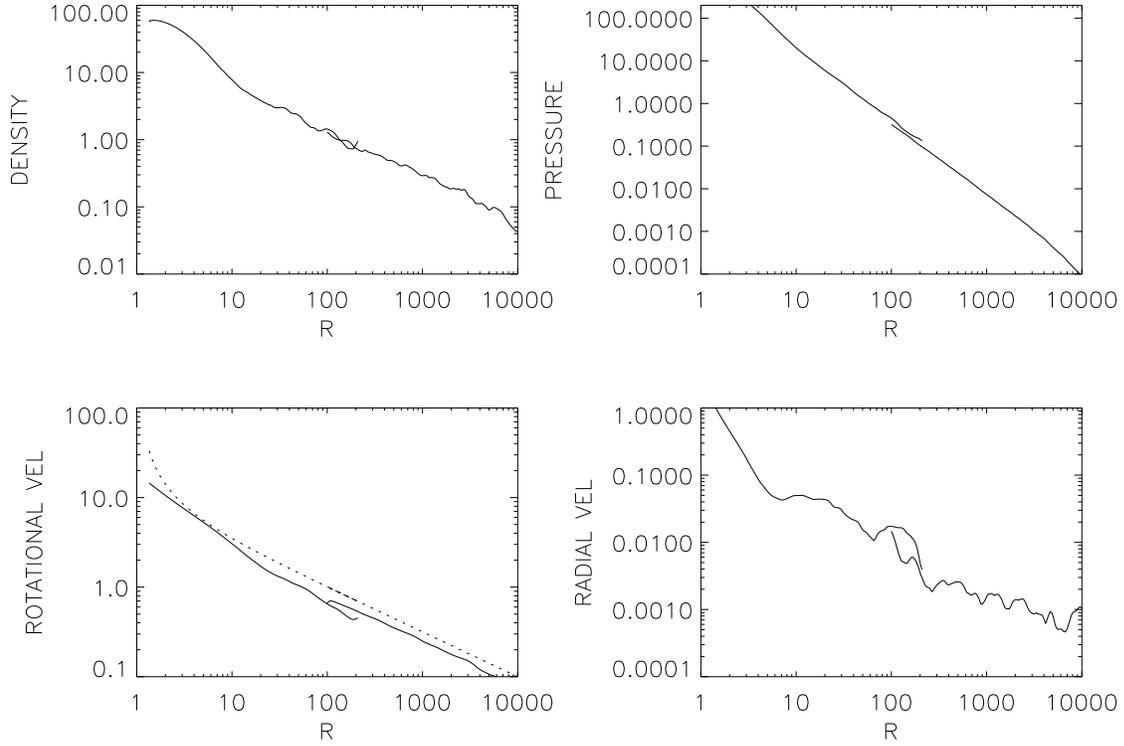}
\caption{Radial structure of the accretion flow (two zones) of Model
B. All quantities have been averaged over time and the polar angle
between $\theta=84^{\circ} $ and $\theta=96^{\circ}$. The dashed line in the plot of rotational velocity $v_{\phi}$ denotes Keplerian rotation at the equator. The density
and pressure profiles are slightly steeper than those in Model
A.}\label{fig:radialstructure}
\end{figure*}

\begin{figure*}
\hspace{1.5cm}\includegraphics[width=15cm]{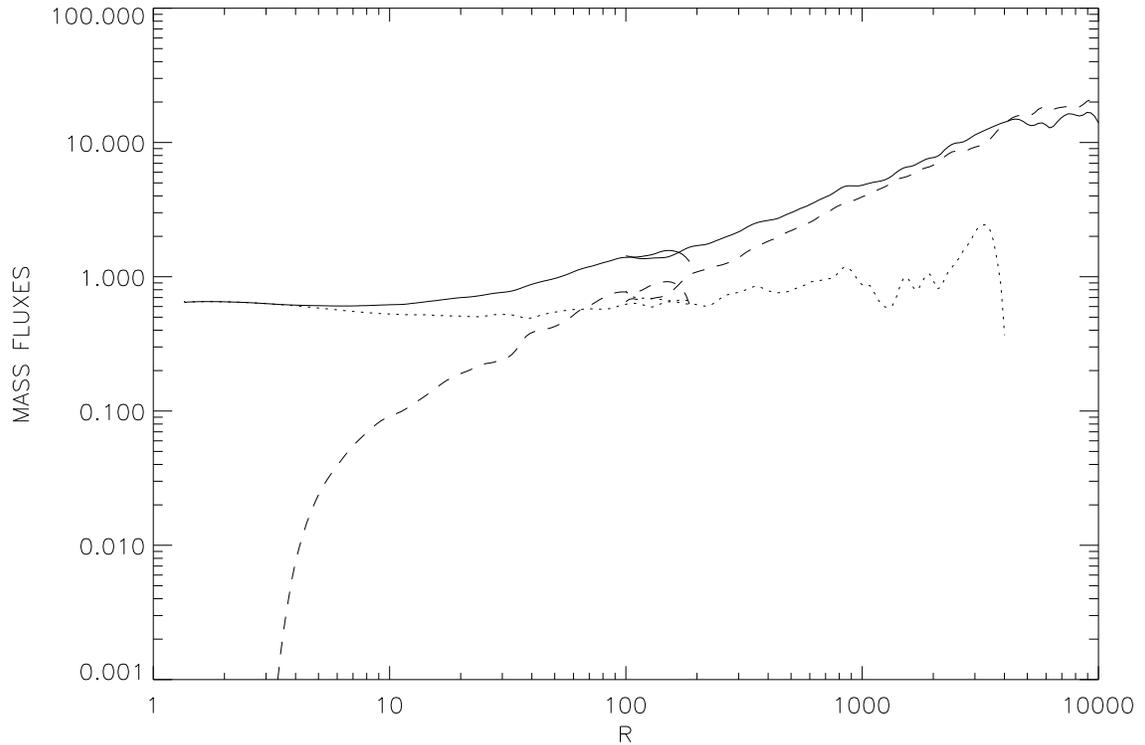}
\caption{Time-averaged and angle integrated mass accretion rate of
Model B. The solid, dashed, and dotted lines are for the mass inflow
rate $\dot{M}_{\rm in}$, outflow rate $\dot{M}_{\rm out}$, and the
net rate $\dot{M}_{\rm acc}$,
respectively.}\label{fig:inflowoutflowB}
\end{figure*}

We simulate Model B to examine the effect of varying the amplitude
of the shear stress. The viscosity coefficient in Model B is 10
times larger than Model A.  Igumenshchev \& Abramowicz (1999) and
SPB99 have already investigated this issue. They found that when the
stress becomes stronger, the convective instability becomes
weaker, in the sense that the radial profile of inflow rate becomes
flatter. But they did not present the qualitative results of the
profiles of density and inflow rate when the stress is increased. In the extreme case, Igumenshchev \& Abramowicz (1999) found that when $\alpha\ga 0.1$, the flow becomes almost laminar, and a bipolar outflow structure very close to the rotation axis, $\sim 8 r_s$, is produced.

In the present work, for the aim of comparing with observations (see
\S4 below), we try to give more qualitative results. Figures 3 is similar
to Figure 1, presenting the radial profiles of density, pressure,
rotational velocity, and radial velocity. Figure 4 shows the radial profiles of inflow and outflow rates and the net rate. From Figure 3, we see that the
equatorial density, gas pressure, rotation velocity and radial
velocity can again be well described by a power-law form. The slopes
of the profiles of rotation and radial velocity remain the same with
Model A. The normalization of the radial velocity is 10 times
larger, as expected. Compared to Model A, the profiles of density and pressure
are moderately steeper, while that of the inflow rate becomes
flatter, \be\rho(r) \propto r^{-0.85}, \hspace{1cm}p (r) \propto r^{-1.85},
\ee \be \dot{M}_{\rm in}(r)\propto r^{0.4}.\ee
Again more precisely, the
inflow rate is described by \be\dot{M}_{\rm in}(r) \propto r^{0.54}\label{modelbmdotprofile}\ee from
$\sim 10r_s$ to $10^4 r_s$, while it is almost constant within
$10r_s$. The outflow rate is much steeper. Beyond $r=100$ it can be described by \be\dot{M}_{\rm out}(r)\propto r^{0.73}.\ee Again we see that the relationship of $p=1.5-s$ is reasonably satisfied beyond $10r_s$, as in the case of $\alpha=0.001$.

We have also conducted simulations with $\alpha=0.05$. We found that
convective outflow becomes significantly weaker. The density profile
becomes steeper while the accretion rate profiles becomes flatter.

\subsection{Effect of changing initial conditions}

In Model A and B, a rotating torus is adopted as the initial condition of our simulation. We also study the cases of other initial condition. One is to inject gas from an outer boundary, with the properties of the injected gas, including the temperature, rotation and radial velocity, being determined by the self-similar solution of an ADAF (Narayan \& Yi 1994). Another model is that the initial condition is  expanded from the one dimensional global solution of ADAFs. We find that the radial profiles of inflow rate and density are very similar (Paper II). However, this will not be the case if the angular momentum of the injected gas in the initial condition is very low. In that case, the profile of accretion rate will become flatter and correspondingly density profile steeper. The full discussion of the effect of initial condition and boundary condition will be presented in a subsequent paper (Bu, Yuan \& Wu 2012).

\subsection{The new scaling law of hot accretion flow}

Narayan \& Yi (1995; see also Narayan, Mahadevan \& Quataert 1998)
presented the radial scaling of many quantities of the hot accretion
flow based on the self-similar solution. These solutions are very
useful to estimate approximately the properties of hot accretion
flow. But at that time, the inward decrease of the accretion rate has not been
found and taken into account in the solution.
For the convenience of future use, here we present
the new scaling law. In terms of interpreting observations, the most prominent effect of outflow is on the profile of density. Therefore, compared to Narayan \& Yi (1995), only the scaling of density-related quantities such as the number density, magnetic field
strength, pressure, and viscous dissipation rate, are changed. The scaling of other quantities are also presented for completeness.
In the original scaling of Narayan \& Yi (1995), the parameter
$\beta=0.5$ ($\equiv p_{\rm gas}/(p_{\rm gas}+p_{\rm mag})$, with
$p_{\rm mag}=B^2/8\pi)$. Here we do not specify a value for $\beta$. The new results are,
$$ v\approx -1.1\times 10^{11}\alpha r^{-1/2}~{\rm cm~s^{-1}}$$ $$
\Omega\approx 2.9\times 10^4m^{-1}r^{-3/2}~{\rm s^{-1}}$$
$$c_s^2\approx 1.4\times 10^{20}r^{-1}~{\rm cm}^2~{\rm s}^{-2}$$
$$n_e\approx 6.3\times 10^{19}\alpha^{-1}m^{-1}\dot{m}_{\rm out}r_{\rm
out}^{-3/2}\left(\frac{r}{r_{\rm out}}\right)^{-p}~{\rm cm}^{-3}$$
$$B\approx 6.5\times 10^8 (1-\beta)^{1/2}
\alpha^{-1/2}m^{-1/2}\dot{m}_{\rm out}^{1/2}r_{\rm
out}^{p/2-3/4}r^{-1/2-p/2}~{\rm G}$$
$$p\approx 1.7\times 10^{16}\alpha^{-1}m^{-1}\dot{m}_{\rm out}r_{\rm
out}^{p-3/2}r^{-1-p}~{\rm g~cm^{-1}~s^{-2}}$$
$$q^+\approx 5.0\times 10^{21}m^{-2}\dot{m}_{\rm out}r^{-5/2-p}r_{\rm out}^{p-3/2}~{\rm erg~cm^{-1}~s^{-2}}$$
\be\tau_{\rm es}\approx 24\alpha^{-1}\dot{m}_{\rm out}r^{1-p}r_{\rm
out}^{p-3/2}\ee Same with Narayan \& Yi (1995), all quantities are
written in scaled units, $M=m\msun$, $r$ is radius in unit of $r_s$,
$\dot{M}=\dot{m}\dot{M}_{\rm Edd}$ ($\dot{M}_{\rm Edd}\equiv
10L_{\rm Edd}/c^2)$, $\dot{m}_{\rm out}$ is the mass accretion rate
at the outer boundary $r_{\rm out}$.

\section{Comparison with Previous Works and Observations}

\subsection{Comparison with previous simulations: a common radial density profile?}

SPB99 found that, for the $\alpha$-description (Run K in SPB99) as we adopte in the present work, $\rho (r)\propto r^{-0.5}$. Igumenshchev \& Abramowicz (2000) performed two-dimensional HD simulations, considering a larger range in the parameter space spanned by $\alpha$ and adiabatic index $\gamma$. They found that when $\gamma=5/3$, for both $\alpha=0.01$ and $0.03$ the density profile is roughly the same, i.e., $\rho (r) \propto r^{-0.5}$. This result is approximately consistent with SPB99.  The density profile in these two works is flatter than our result since a Newtonian potential was adopted there. McKinney \& Gammie (2002) obtained $\rho\propto r^{-0.6}$ and $v_r\propto r^{-2}$. But this result strongly suffers from the ``inner boundary effect'', since a pseudo-Newtonian potential was adopted and the above scaling was measured over $1.3 r_s \la r \la 10r_s$.

Many MHD numerical simulations of hot accretion flow have been done
after the pioneer HD works of Igumenshchev \&
Abramowicz (1999, 2000) and SPB99. For two-dimensional simulations, since turbulence and accretion die away because there is
no dynamo action to maintain the poloidal magnetic field, steady
state can not be reached at large radii. This effect, combined with
the inner boundary effect, severely restricts the measurement of the
radial profiles of dynamical quantities such as density. For three-dimensional simulation, it is very time-consuming to simulate a large dynamical range. Perhaps due to these reasons, only a few works have presented radial profiles of density and inflow rate. For example, Stone \& Pringle (2001), Hawley, Balbus \& Stone (2001), and Hawley \& Balbus (2002) did the two and three-dimensional MHD simulation of accretion flow. They all found that the mass accretion rate decreases with decreasing radius, but the quantitative radial profile was not given.

Machida, Matsumoto \& Mineshige (2001) performed three-dimensional
MHD simulation of hot accretion flows, with a toroidal initial magnetic field
configuration. A Newtonian potential was adopted thus their results are not affected by the ``inner boundary effect''.  They found that the radial structure of the accretion flow is very similar to that obtained by HD simulations. Two fits were obtained. In the ``early stage'' of the simulation they found that the radial profiles of the density and radial velocity can be described by $\rho\propto r^{-0.5}$ and $v_r\propto r^{-1.5}$. At the ``late stage'' of their simulation, they found moderately different results, $\rho \propto r^{-0.8}$ and $v_r\propto r^{-1.3}$. While the profile of density is very similar to our result, their profile of radial velocity is much steeper. This may be because that the averaging approach adopted in their work is different from ours. They average the quantity over the whole $\theta$ angles (eqs. [1] and [2] in Machida, Matsumoto \& Mineshige 2001); while our average is constrained to $84^{\circ}\le \theta\le 96^{\circ}$. This approach is expected to produce a quite different result only to the radial velocity, since the radial velocity away from the equatorial plane is much larger than that close to $\theta\sim 90^{\circ}$ while density is on the opposite.  Igumenshchev, Narayan \& Abramowicz (2003) also presented three-dimensional MHD simulation. They adopted Paczy\'nsky \& Wiita potential and  two types of initial magnetic field configurations. Only in the case of a toroidal initial configuration the density profile was given, which is $\rho\propto
r^{-1}$. This profile is steeper than that of Machida, Matsumoto \& Mineshige (2001). The discrepancy should not because of the inner boundary effect, since the measurement is over  $7r_s \la r \la 100r_s$. We speculate that one reason may be that a different potential was adopted. Another reason may be associated with the strong fluctuation of the radial velocity in this radial range (see Fig. 11 in Igumenshchev, Narayan \& Abramowicz 2003).

In the above-mentioned MHD simulations, the magnetic field is weak. Recently attention has been paid to accretion flows with accumulated strong magnetic flux, i.e., ``magnetically arrested disk'' or ``magnetically choked accretion flows'' (Narayan, Igumenshchev \& Abramowicz 2003; Pen et al. 2003; Pang et al. 2011; McKinney, Tchekhovskoy \& Blandford 2012). Pen et al. (2003) constructed three-dimensional MHD simulation with huge number ($1400^3$) of grid zones. They emphasized that the axial and reflection
symmetries usually adopted in many simulations are not adopted in
their work. The input magnetic field is different from the usual way. It it an admixture of random field loops and large loops that thread the whole simulation box. The density profile they measured is $\rho\propto r^{-0.72}$ over the radial range of $0.03r_B\sim 0.3r_B$, with $r_B$ is the Bondi radius. The large distance of the inner boundary from the black hole implies that this result does not suffer from the ``inner boundary effect''. This slope is well between the results of our Model A and Model B.

In another three-dimensional MHD simulation  by Pang et al. (2011), special attention was paid to the radial density (or equivalently accretion rate) profile of the accretion flow. A Newtonian potential was adopted in their work. The outer boundary is quite large, almost ten times larger than the Bondi radius. They conducted a numerical survey of parameter space, namely the magnitude of the ambient magnetic field, the radial dynamical range within Bondi radius, and the resolution of the Bondi scale. The found that the mass accretion rate at the inner boundary $r_{\rm in}$ can be well described by the following power-law form, \be \frac{\dot{M}}{\dot{M}_{\rm Bondi}}=\left(\frac{r_{\rm in}}{r_{\rm B}}\right)^{3/2-k_{\rm eff}},\ee with the mostly favored value of $k_{\rm eff}\approx 1$. Here $\dot{M}_{\rm Bondi}$ is the accretion rate at Bondi radius $r_{\rm B}$ and $r_{\rm in}\sim 0.03r_{\rm B}$.  Again like the case of Pen et al. (2003), the rather large $r_{\rm in}$ ensures that the obtained profile is not affected by the inner boundary effect. This result is in good agreement with our Model B (eq.[\ref{modelbmdotprofile}]).

Mckinney, Tchekhovskoy, \& Blandford (2012) simulated three-dimensional general relativistic MHD accretion flow with different thickness and two types of initial magnetic field configuration, namely poloidal and toroidal. Density profiles haven been obtained for ``thick'' flows, which is $\rho\propto r^{-0.6}$, same for both configurations. For ``thin'' flows, it is $\rho \propto r^{-0.7}$. However, the measurement is not relevant to us since it is over $6r_s \la r \la 15r_s$, where the inner boundary effect is very strong.

Narayan et al. (2012) most recently presented two long-duration GRMHD simulations of a hot accretion flow, in which the accumulation of magnetic flux around the black hole does and does not occur (``MAD'' and ``SANE'' models), respectively. The steady state is reached up to $\sim 50 r_s$. After taking into account the ``inner boundary effect'' and ``$\alpha (r)$ effect'',  they found $v_r\propto r^{-0.5}$ when $r\ga 10r_s$. Regarding the density profile, they claimed  that its radial scaling is such that $\dot{M}(r)\sim $ constant is roughly satisfied at small radii. Note that this is not in conflict with our results. As we have stated before and shown by Figs. 2\&4, $\dot{M}(r)\approx {\rm constant}$ when $r\la 10r_s$.  Outside $\sim 10r_s$ the radial inflow rate profiles for the ``MAD'' and ``SANE'' models are $\dot{M}(r)_{\rm in}\propto r^{0.4}$ and $\propto r^{0.65}$, respectively (Ramesh Narayan, private communications). The result of the ``SANE'' model is in good consistency with our Model A (eq.[\ref{inflowprofile2}]) while the result of ``MAD'' model is similar to our Model B (eq.[\ref{modelbmdotprofile}]). From the above-mentioned works we note that the inflow rate profile of accretion flow with accumulated magnetic flux seems to be somewhat flatter than that without accumulation. This may be because that when there is accumulation of magnetic flux, the ``effective $\alpha$'' is larger, i.e., ``MAD'' and ``SANE'' correspond to our Model B and A, respectively.

Overall, almost all current numerical simulations give a similar radial density profile, with $\rho\propto r^{-(0.5-1)}$. The results depend weakly on the presence or absence of magnetic field (HD or MHD), viscous parameter $\alpha$, strength of the magnetic field (weak or strong), the initial configuration of the magnetic field (toroidal or poloidal or tangled), and the dimension of calculation (two or three dimension). If this is true, an interesting question is then: what causes such a result? To answer this question, one first needs to understand what causes the inward decrease of the mass accretion rate. This question will be studied in Paper II. Here we briefly summarize the results. Two scenarios have been proposed to explain this result. One is the convection-dominated accretion flow (CDAF) model (Narayan et al. 2000; Quataert \& Gruzinov 2000a). This model proposes that both hydro and MHD accretion flows are convectively unstable. In this case the inward decrease of accretion rate is because that the fluid circulates in the convective eddies. In contrast to the CDAF model, the adiabatic inflow-outflow solution (ADIOS) model (Blandford \& Begelman 1999, 2004; Begelman 2012) suggests that the inward decrease of accretion rate is because of mass loss in the outflow. Our analysis  in Paper II favors the latter. Moreover, we find that outflow is driven by buoyant force and magnetic centrifugal force in HD and MHD accretion flows, respectively.

Then it is an interesting question why different physical mechanisms cause the roughly same profiles of accretion rate and density? Begelman (2012) gives an answer to this question. In the early version of the ADIOS model (Blandford \& Begelman 1999, 2004), there are inflowing and outflowing zones with equal but opposite mass fluxes that vary with radius following a power-law form, $\dot{M}\propto r^s$. The net rate, which is the sum of inflow and outflow rates, is very small. They applied the conservation laws of hydrodynamics to the inflow zone and obtained relationships among the parameters such as the specific angular momentum, effective binding energy, and the value of $s$. In this model, the value of $s$ is not determined but only constrained to be in the range $0\le s \le 1$. One advantage of the ADIOS model, as claimed by the authors, is that although it was based on HD analysis, the approach can be fully applied to MHD accretion flows as well. But the disadvantage is that the early ADIOS model is intrinsically one-dimensional, the outflow zone is laminar and their streamlines are self-consistently patched onto the inflow zone at each radius. Begelman (2012) recently developed this model. The main improvement is that now the inflow and outflow zones are treated on an equal footing,  and the outflow is permitted to be as turbulent as inflow rather than a laminar flow.  He found that the viable range of $s$ is now narrowed down from $0\le s \le 1$ to $s\approx 1$. The main reason for the regulation of the solution is an introduction of a conserved outward flux of energy through the flow, which was ignored in Blandford \& Begelman (1999, 2004). This work therefore well explains why various simulations obtain very similar profiles of accretion rate and density. Because of this reason, we have confidence that our HD simulation results should be similar to the more realistic MHD simulation results, and further can be compared to observations. Of course, we note that the quantitative result of the value of $s$ obtained in Begelman (2012) is moderately larger than that obtained in simulations. In Model A, when $r\ga 100 r_s$ where the ``inner boundary effect'' is the weakest thus more appropriate to be compared with Begelman (2012), we find $s\sim 0.75$.

\subsection{Comparisons with observations}
\subsubsection{Sgr A*}

Because of its proximity, the supermassive black hole located at the
center of our Galaxy is regarded as the unique laboratory for the
study of black hole accretion (see Yuan 2011 for a review). Abundant of data has been obtained, which puts the most
strict constrains to the theory of black hole accretion. Among them
is the constrain on the mass accretion rate at both the Bondi radius
and the region close to the black hole horizon. The high spacial
resolution of {\em Chandra} can well resolve the Bondi radius and
infer the gas density and temperature there (Baganoff et al. 2003).
The values of the Bondi radius and the Bondi accretion rate are then well determined,
$r_{\rm Bondi}\approx 10^5 r_s$ and $\dot{M}_{\rm Bondi}\approx
10^{-5}\mpy$, respectively. Although the value of mass accretion rate is obtained by the simple analytical Bondi theory, it was found to be in good consistency with the detailed three-dimensional numerical simulations focusing on the fueling of the supermassive black hole in Sgr A* of Cuadra et al. (2006). In fact, the numerical simulations obtained in this work is $\dot{M}\approx 3 \times 10^{-6} \mpy$, only a factor of 3 lower than $\dot{M}_{\rm Bondi}$. This factor is likely because of the inclusion of angular momentum of the accretion flow in the simulation.

The accretion rate at the innermost region of the accretion flow, on the other hand, is strongly constrained by the radio polarization observations (e.g., Quataert \& Gruzinov 2000b; Macquart et al. 2006; Marrone et al. 2007). The detected high level of linear
polarization at frequencies higher than $\sim 150 {\rm GHz}$ sets an
upper limit to the rotation measure, which then requires
that the accretion rate at the innermost region of the accretion flow must be in the range between $2\times 10^{-7}\mpy$ and $2\times 10^{-9}\mpy$, depending
on the assumed configuration of the magnetic field. This implies
that more than $99\%$ of the gas captured at the Bondi radius must be lost and does not fall onto the  black hole. Correspondingly, the density profile of the accretion flow is
significantly flatter than the prediction of $\rho\propto r^{-3/2}$. The detailed modeling to Sgr A* presented in Yuan, Quataert \& Narayan (2003) has shown that the radial profiles of accretion rate and density are described by $\dot{M}\propto r^{0.3}$ and $\rho\propto r^{-1}$.

This density profile is moderately steeper than
that obtained in our simulation. This is likely because that the specific angular momentum of the accretion flow at the Bondi radius is significantly sub-Keplerian, $\sim 0.3$ times Keplerian, as the numerical simulation has indicated (Cuadra et al. 2006). In this case, our study (Bu, Yuan \& Wu 2012) indicates that the outflow becomes significantly weaker, and the density profile correspondingly steeper.

\subsubsection{NGC~3115}

NGC~3115 is an low-luminosity AGN. The mass of the black hole in its center is $M = (1-2)\times 10^9\msun$. The source is very dim. {\em Chandra} observations only
present an upper limit of $10^{38}\ergs\sim 10^{-9}L_{\rm Edd}$
(Wong et al. 2011 and references therein). Therefore the accretion
mode in this source must be an ADAF. At a distance of 9.7 Mpc, NGC~3115 is the nearest $> 10^9\msun$ black hole. Therefore, the angular size of the accretion flow in this source is very large, which is very helpful for us to directly detect the accretion flow. Wong et al. (2011) recently conducted {\em Chandra} observation to this
source and determined that the Bondi radius is about
$4^{\arcsec}-5^{\arcsec}$. The most prominent result is that for the
first time, they have resolved the accretion flow within the Bondi
radius. They found that the temperature is rising toward the black
hole as expected in all accretion flow models. But more importantly, the
radial density profile of the accretion flow within $4^{\arcsec}$ is
found to be well described by a power-law form, \be \rho(r) \propto
r^{-1.03^{+0.23}_{-0.21}}. \ee This result is similar to Sgr A* and is in reasonable agreement with our numerical simulations.

\section{Summary}

One of the most important finding of both HD and MHD numerical simulations of hot accretion flow in the recent years is that the mass accretion rate decreases with decreasing radius. Correspondingly, the density profile becomes flatter compared to the previous prediction when the accretion rate is constant of radius. One main problem with previous simulation is that because of technical difficulty the radial dynamical range in simulations is usually very limited, typically spanning less than two orders of magnitude. The boundary condition effect makes the radial range over which we can reliably measure the profile of physical quantities even smaller. The previous simulation results are therefore somewhat suspectable. In this paper we adopt a ``two-zone'' approach to simulate an axisymmetric accretion flow, extending the dynamical range to over four orders of magnitude, i.e, from $\sim r_s$ to $40000r_s$. We confirm previous results that the profiles of inflow rate and density can be well described by power-law forms. Within $10r_s$, $\dot{M}_{\rm in}(r)\sim const.$. Beyond $10r_s$, the power-law slopes are a function of the viscous parameter $\alpha$. For $\alpha=0.001$, they are described by $\rho(r)\propto r^{-0.65}$ and $\dot{M}_{\rm in}(r)\propto r^{0.65}$. For $\alpha=0.01$, the results are $\rho(r)\propto r^{-0.85}$ and $\dot{M}_{\rm in}(r)\propto r^{0.54}$.

We also combine from literature all available numerical simulations which have presented the radial profile of density, both two-dimensional and three-dimensional, HD and MHD. We find that all these simulations give somewhat similar results, $\rho \propto r^{-(0.5-1)}$, and this is also consistent with our results. The diversity of the power-law index seems to come from the differences of the value of $\alpha$, the gravitational potential of the black hole, the initial condition of the simulation, { and the strength of the magnetic field. The rough consistency among} various simulations can be explained by the most recent ADIOS (adiabatic inflow-outflow solution) model (Begelman 2012). In this work it was found that after considering both the inflow and outflow zones at the equal footing and  a conserved outward energy flux, the value of $s$ is well constrained to be in a narrow range.

The radial profiles of accretion rate and density obtained by numerical simulations are in good agreement with observations. In the case of Sgr A*, detailed modeling to the
multi-waveband spectrum find that $\dot{M}(r)\propto r^{0.3}$ and $\rho\propto r^{-1}$. In the case of NGC~3115, for the first time,  {\em Chandra} observations resolve the
accretion flow within Bondi radius, and find $\rho\propto r^{-1.03^{+0.23}_{-0.21}}$.

\section{ACKNOWLEDGMENTS}

We are grateful to Ramesh Narayan for many valuable discussions and constructive comments. We also thank Jerry Ostriker and Jim Stone for stimulating discussions on the two-zone approach adopted in this paper, { and Jon McKinney for the useful comments}. The hospitality of Princeton University and Copernicus Astronomical Center is acknowledged where part of the work was done. This work was supported in part by the National Basic Research Program of China (973 Program 2009CB824800), Natural Science Foundation of China (grants 10833002, 10825314, 11121062, 11103059, and 11133005), and the CAS/SAFEA International Partnership Program for Creative Research Teams. The simulations were carried out at Shanghai Supercomputer Center.

\end{document}